\documentclass[amsmath,amssymb,aps,pre,reprint,superscriptaddress]{revtex4-2}
\usepackage{graphicx}% Include figure files
\usepackage{dcolumn}% Align table columns on decimal point
\usepackage{bm}% bold math
\newcommand\Imag{\mbox{Im}} % cf plain TeX's \Im

\begin{document}

\title{Effects of finite non-gaussianity on evolution of a random wind wave field}
\author{S.Y. Annenkov}
\affiliation{School of Computing and Mathematics, Keele University, Keele ST5 5BG UK}
\affiliation{Shirshov Institute of Oceanology, Russian Academy of Sciences}
\author{V.I. Shrira}
\affiliation{School of Computing and Mathematics, Keele University, Keele ST5 5BG UK}
\date{\today}
\begin{abstract}
We examine long-term evolution of a random wind wave field
generated by constant forcing, by comparing  numerical
simulations  of the kinetic equation and direct numerical
simulations (DNS) of the dynamical equations. While
integral characteristics of spectra are in reasonably
good agreement, the spectral shapes differ considerably at
large times, the DNS spectral shape being in much better
agreement with field observations. Varying the number of
resonant and approximately resonant wave interactions in
the DNS numerical scheme, we show that when the ratio of
nonlinear and linear parts of the Hamiltonian tends to
zero, the DNS spectral shape approaches the shape predicted by
the kinetic equation. We attribute the discrepancies between
the kinetic equation modelling, on one side, and the DNS and
observations, on the other, to the neglect of
non-gaussianity in the derivation of the kinetic equation.
\end{abstract}

\maketitle

\emph{Introduction.}---Long-term nonlinear evolution of
random wave fields is described by wave turbulence theory,
which links ensemble averaged  quantities of the field to
spectral density $n_{\mathbf k} = n({\mathbf k}, t)$, a
function of wavevector ${\mathbf k}$ and time $t$
\cite{ZLF}. This quantity, proportionate to the Fourier
transform of two-point averages, is called the particle
number, power or waveaction density in various contexts. In
many cases, this quantity develops an inverse cascade
towards large scales, leading to emergence of long waves
due to  resonant interactions of shorter waves. Within the
wave turbulence theory, the process is described by the
kinetic equation (KE), which expresses time derivatives of
the density in terms of this density only
\cite{NewellRumpf2013, ZLF}. Analytical and numerical
solutions of this equation form the core of our
understanding of random wave fields evolution on large
timescales.

By far the most studied example of this evolution is 
provided by oceanic wind waves. The corresponding KE is
known as the Hasselmann equation \cite{Hass62}, and is
routinely simulated numerically for operational wave
forecasting. From the theoretical perspective, the wind
wave example stands out due to the continuous world wide
testing of the forecasting against observations, which
gives a unique chance to verify the assumptions underlying
the wave turbulence theory.

While there is a consensus that the KE does capture main
features of wave field evolution
\cite{Komen,Toba72,Hasselmann76,ResioVincent,Zakh2015}, 
there are major discrepancies between
the KE based predictions and observations. The inherent
property of the KE is the homogeneity of the wave
interaction term as a function of spectral energy density
\cite{Zakh2015}. This property leads to self-similarity of
the solutions in a wide range of wave generation conditions
\cite{Badulin2005} and allows to formulate the basic laws
of wind-wave growth independently of wind speed
\cite{Zakh2015}. In the idealized situation when a wave
field is generated by constant wind, the KE predicts that
the solution tends to a permanent self-similar shape, with
a characteristic enhanced peak and steep nearly straight
spectral front, which evolves towards large scales
following the known asymptotic laws (e.g.\
\cite{Badulin2005}). As long as the consideration is
limited to integral characteristics of a wave field (such
as significant wave height, total energy, frequency of the
spectral peak), the self-similar picture of wind wave
development predicted by the KE is generally supported
experimentally \cite{Toba72,Hasselmann76, Zakh2015}.
However, a closer look at spectral shapes reveals a major
discrepancy:  the observed spectral shapes of young and
mature sea states essentially differ (JONSWAP and
Pierson-Moskowitz spectra respectively) \cite{Holthuijsen}.
As waves mature, a decrease in the spectral peakedness is
observed \cite{BabaninSoloviev,LongResio,ResioVincent,
RomeroMelville2010}. 
The parametrization of spectral shape of fully-developed sea,
proposed in \cite{PM64}, confirmed in later reanalysis
\cite{Alves} and at present widely accepted to be well
assured statistically \cite{BabaninSoloviev}, is different
from the shape of the self-similar KE solutions, having a
more rounded spectral front and peak with no pronounced
enhancement. Evolution of the spectral shape with fetch
with a decrease of peakedness and a continuous transition
to the Pierson-Moskowitz spectrum has been described in a
number of measurement-based studies
\cite{BabaninSoloviev,LongResio,ResioVincent,
RomeroMelville2010}. At this
stage of wave field development, the spectral peak is no
longer under direct wind forcing, and the probability of
breaking events for dominant waves is  low
\cite{Banner,Alves}, so that the shape of the peak should
be determined primarily by nonlinear interactions.
Therefore, the fact that the KE is unable to reproduce this
spectral shape represents  a major fundamental challenge.
%theoretical explanation.

Until recently, the KE as the model of long-term wind wave
field evolution did not have an alternative (a
generalization proposed by \cite{ASh2006}, although useful,
in the large time limit tends to the KE and, hence, does
not resolve the contradiction). This situation was changed
when a direct numerical simulation (DNS) algorithm capable
of long-term simulations of random wave field was proposed
\cite{ASh2013,AShWTBook,ASh2018}. The algorithm is based on
the Zakharov equation, from which the KE is derived, and
performs simulations with ensemble averaging without any
statistical assumptions. DNS simulations without wind
showed, in particular, that in contrast to the KE predictions
the DNS spectra have less steep, more rounded spectral
front and considerably wider
and lower peak %, and demonstrated a departure from the
%strict kinetic scaling of growth rates, predicted by the KE
\cite{ASh2018}. At the same time, no apparent reason to
question the basic assumptions underlying the wave
turbulence theory (e.g.\ smallness of non-gaussianity, wide
separation of timescales) has been found, and an
explanation of the discrepancies remained outstanding.

In this work, we identify the origin of the discrepancies
as the neglect of small but finite non-gaussianity in the
derivation of the KE itself, rather than in the underlying
theory. In order to get a closed equation in terms of
$n_{\mathbf k}$, one has to express the six-point
correlator in terms of two-point ones only, neglecting the
four-point cumulants in the expansion \cite{Zakharov99}.
This approximation, equivalent to the assumption of random
initial phases \cite{OnoratoDematteis}, leads to the
aforementioned homogeneity property, absent in the
primitive equations, and contradicts the established
picture of weak turbulence, according to which the wave
field evolution is due to nonlinear regeneration of
non-gaussianity \cite{NewellNazarenkoBiven}. Thus, the KE
takes into account finite nonlinearity (the asymptotic
expansion in powers of nonlinearity retains cubic terms),
but assumes infinitesimal non-gaussianity. Crucially,
according to the KE the spectral shape %reached in the
%large-time limit of evolution under quasi-stationary
%conditions
is the same for all levels of nonlinearity, including
infinitesimal nonlinearity. In this sense, the neglect of
finite non-gaussianity does not allow to properly
capture the effects of finite nonlinearity either.

To consider the effects of finite non-gaussianity on the
long-term evolution of a random wave field, we examine by
DNS the evolution of a wave field generated by constant
wind, and compare it with the KE results. In both models,
the evolution tends to self-similarity, and integral
characteristics of the spectra are close to each other.
However, the shape of the DNS spectra is quite different,
with a lower, less pronounced peak. Introducing an integral
characteristics of the non-gaussianity linked to the
coarse-graining parameter of the numerical scheme, we find
the wave spectrum dependence on non-gaussianity, and
demonstrate that the DNS spectral shape converges to the KE
one when the non-gaussianity tends to zero. This enables
us to attribute the origin of the discrepancy to the
neglect of non-gaussianity in the KE derivation.

\emph{Theoretical background and numerical methods.}---We
consider  gravity waves on the surface of deep fluid
governed by the Zakharov equation \cite{ZLF}
\begin{equation}
\label{ZakhEq}
{\rm i}\displaystyle\frac{\partial b_0}{\partial t}  =
  \omega _0 b_0+ \displaystyle\int T_{0123}b_1^{*}b_2b_3
  \delta _{0+1-2-3}\,{\rm d}{\bf k}_{123}.
\end{equation}
Here, $b({\bf k})$ is a canonical complex variable in
Fourier space, ${\bf k}$ is the wavevector, $k=|{\bf k}|$,
$\omega ({\bf k})=\left( gk\right)^{1/2}$ is the linear
dispersion relation. The compact notation used designates
arguments by indices, e.g., $T_{0123}=T({\bf k},{\bf
k}_1, {\bf k}_2,{\bf k}_3)$, $\delta _{0+1-2-3}=\delta
({\bf k}+{\bf k}_1-{\bf k}_2-{\bf k}_3)$, asterisk means
complex conjugation, and $t$ is time.

For wave fields with the $2 \leftrightarrow 2$ type
dominant resonant process    (\ref{ZakhEq}) is often taken
as the primitive equation (thus, higher-order resonances
are neglected). Then the
statistical description of a wave field can be obtained in
terms of correlators of $b({\bf k},t)$ as \cite{ASh2006}
\begin{equation}
\label{j1}
\displaystyle\frac{\partial n_0 }{\partial t} = 2\Imag \int
T_{0123} J^{(1)}_{0123} \delta_{0+1-2-3}\,{\rm d}{\bf
k}_{123}.% + 2\gamma_0n_0.
\end{equation}
where $n_0$ is the second-order correlator, $\langle
b_0^*b_1 \rangle=n_0\delta_{0-1}$,
$J^{(1)}_{0123}$ is the four-point cumulant.
In the next order (\cite{Zakharov99}),
\begin{widetext}
\begin{eqnarray}
\label{ddt}
\left(\displaystyle\frac{\partial}{\partial t} - \mathrm{i}\Delta\omega\right)J^{(1)}_{0123}
  & = & 2 \mathrm{i} \int \Bigl \{ T_{0456}\delta_{0+4-5-6}I_{156234}
   \\ \nonumber
  && + T_{1456}\delta_{1+4-5-6}I_{056234} - T_{2456}\delta_{2+4-5-6}I_{014356} - T_{3456}\delta_{3+4-5-6}I_{014256}\Bigr\}\,{\rm d}{\bf k}_{456},
\end{eqnarray}
\end{widetext}
where $\Delta \omega = \omega_0 + \omega_1 - \omega_2 -
\omega_3$, and $I_{012345}$ is the six-point correlator.
%Equations (\ref{j1}-\ref{ddt}) are exact, and s
Since the six-point cumulant is neglected, $I_{012345}$ is
reduced to a lengthy expression containing pair correlators
and four-point cumulants. Thus, the system of equations for
two- and four-point correlators is closed. However, to
obtain the closed equation for $n_\mathbf{k}$, we must
neglect all four-point cumulants in the expansion for
$I_{012345}$, retaining only the leading-order term. Such a
reduction is equivalent to the gaussianity assumption. The
resulting KE is
\begin{equation}
\label{Kin4}
  \frac{\partial n_0}{\partial t} =
  4\pi \int T_{0123}^2 f_{0123} \delta_{0+1-2-3}
    \delta(\Delta \omega)\,{\rm d}{\bf k}_{123},% +2\Imag
%\omega _0n_0,
\end{equation}
where $f_{0123} = n_2 n_3 (n_0 + n_1) - n_0 n_1 (n_2 +
n_3)$. Equivalently, the KE can be derived  by assuming
initial random phases \cite{OnoratoDematteis}.

\begin{figure} % figure 1
\includegraphics[width=8.6cm]{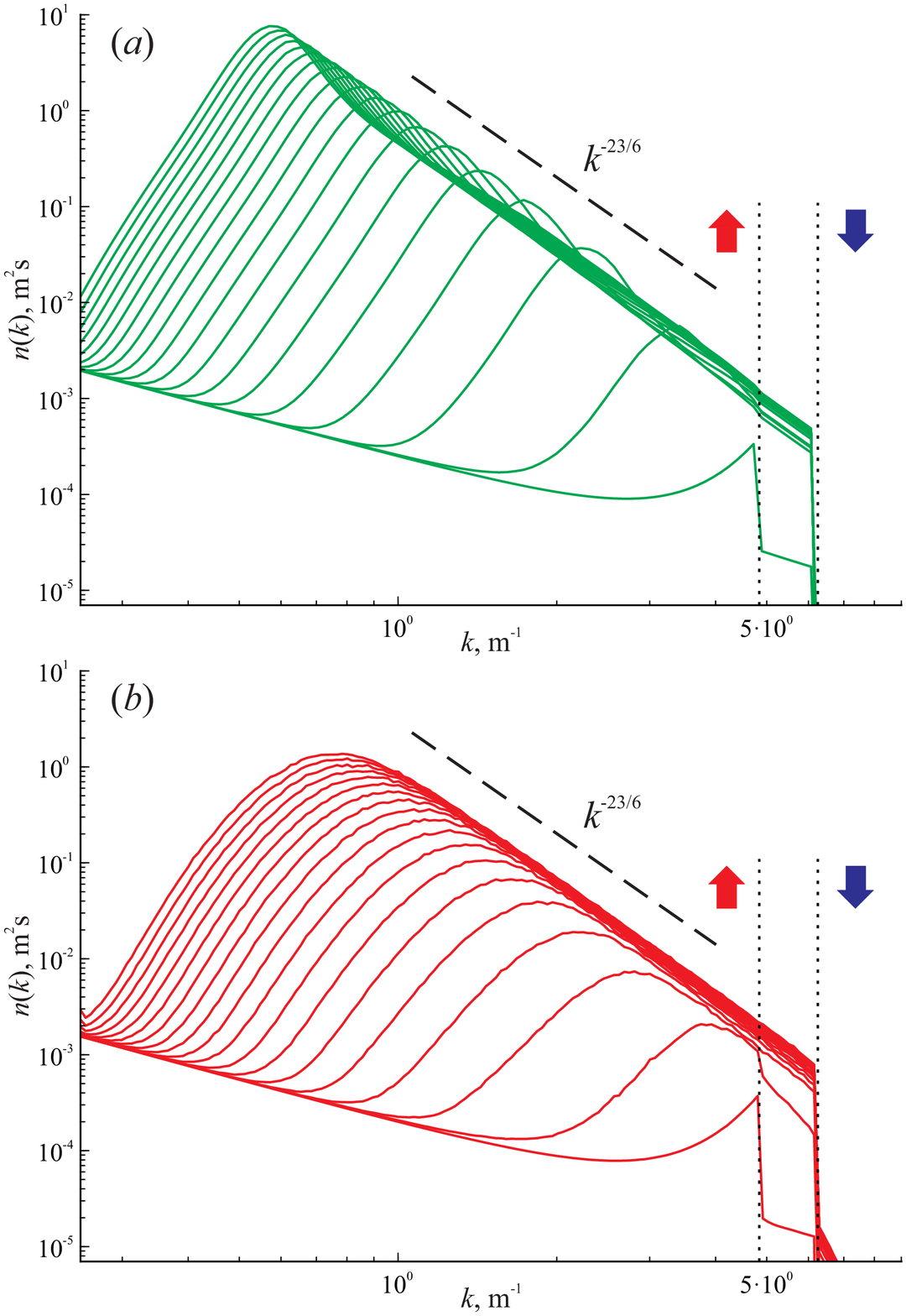}
\caption{Development of wave action spectrum $n(k)$ under
the action of constant wind $U = 2 c_1$, shown in steps of
approximately $300 \tau_1$, where $c_1$ and $\tau_1$ are
characteristic phase speed and period, corresponding to
$k=1$. (\textit{a}) KE, WRT algorithm (\textit{b}) DNS,
with $\lambda_k = 0.03$ on $161\times 71$ grid. Wavenumber
ranges of forcing and dissipation are indicated by arrows}
\end{figure}   % figure 1

The neglect of weak non-gaussianity in the derivation of
the KE makes its right-hand side  a homogeneous function of
$n(\mathbf{k})$ \cite{Zakh2015}. Basically, this means that
the shape of the solutions for spectra provided by the KE
corresponds to the case of infinitesimal amplitude. Thus
the role of finite amplitude effects remains unknown. As
the first attempt to study these effects, we compare
numerical simulations based on the KE and the Zakharov
equation (\ref{ZakhEq}).  The Zakharov equation is
simulated using the original algorithm, described in
\cite{ASh2018}. All parameters of the algorithm are the
same as in \cite{ASh2018}, except for three modifications.
First, in contrast to the simulations of swell in
\cite{ASh2018}, we add wind forcing according to
\cite{Shemdin}, for all $k < 4.84$. Second, to accommodate
the wider angular distribution of wind wave spectra, we
increase the range of angles, retaining the same angular
resolution. Thus, the computational grid contains 161
logarithmically spaced wavenumbers in the range $0.25 \le k
\le 9$ and 71 angles in the range $-7\pi/9\le \theta \le
7\pi/9$. Third, a DNS algorithm with wind forcing can only
be functional if a certain parameterization of wave
breaking is employed. Here we do not attempt to model the
physical process of wave breaking, but intermittency in
individual realisations makes it necessary to limit the
growth of some wind amplified harmonics. To this end the
following empirical rule is introduced: if non-dimensional
value $\varepsilon_k = 0.5\sqrt{2 \omega n_{\mathbf
k}/g}/\pi k$ exceeds $\varepsilon_c$, where $n_{\mathbf k}$
is the discrete wave action of a harmonic under forcing,
then the forcing is changed to damping until $\varepsilon_k
\le 0.1 \varepsilon_c$, when it is resumed. For the
simulations with the $161\times 71$ grid, $\varepsilon_c$
is set to $0.01$, resulting in a small number of ``breaking
events'', usually between 0 and 5 for each realisation at
every timestep. For the KE simulations, we use the standard
WRT algorithm, provided by G. van Vledder, and $101 \times
51$ grid with the same range of wavenumbers and full circle
of directions. Wind forcing is identical for both models,
dissipation is applied to $k \ge 6.25$.

\emph{Results.}---First, we perform simulations of the wave
field development from zero initial condition under forcing
by constant wind with speed $U = 2 c_1$, where $c_1$ is the
phase speed corresponding to $k = 1$, close to the peak
wavenumber at the end of the evolution. The DNS simulation
is performed with averaging over 100 realizations.
Development of the wave action spectrum $n(k)$ with time
for both numerical methods is shown in figure 1. Comparison
of the panels shows that both solutions tend to
self-similar behaviour, and  that the asymptotics of
duration-limited wave growth under constant wave action
flux, known from the analysis of the KE \cite{Badulin2005},
are respected in both cases. However, the spectral shapes,
which are initially close, differ considerably at later
stages of the evolution. Although the spectral slope in
both cases corresponds to the theoretical value for the
inverse cascade of wave action $k^{-23/6}$
\cite{Zakharov99}, the DNS spectra have less steep, more
rounded spectral front and considerably wider and lower
peak. Similar differences were reported earlier for
simulations of wave evolution without wind forcing
\cite{ASh2018}. In the present case, the KE
evolution also demonstrates a slightly %but noticeably
faster downshift rate.

Due to the homogeneity property of the KE, the self-similar
spectral shape is the same for all levels of nonlinearity,
and the downshift rate has a simple scaling law
\cite{Zakh2015}. If the discrepancies between the KE and
the DNS are due to the effects of non-gaussianity
unaccounted for by the KE, they are expected to decrease at
lower levels of nonlinearity. Simulations with lower wind
forcing indeed show that the difference in downshift rate
decreases, but the difference in spectral shapes persists.
Our aim to compare the DNS and the KE evolution in the
small nonlinearity limit cannot be done by simply
decreasing the wind speed, since wind-generated waves
always have a certain finite steepness. %, which decreases
%very slowly during evolution.

\begin{figure} % figure 2
\includegraphics[width=8.6cm]{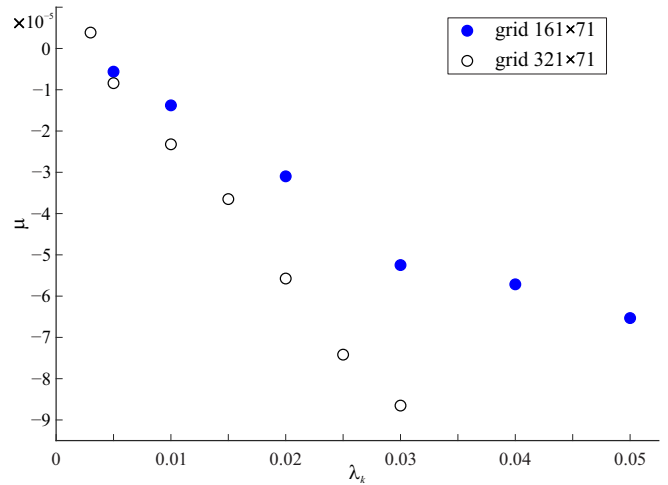}
\caption{Non-gaussianity as measured by $\mu$, the ratio of
statistically averaged nonlinear and linear energy as
function of the coarse-graining parameter $\lambda_k$.
$\mu$ is averaged over 600 characteristic periods of the
final stage of evolution before the spectral peak reaches
$k_p = 1$. }
\end{figure}   % figure 2

Here we use another approach, which is helped by the particular
design of the DNS algorithm. The algorithm is based on the
idea of coarse-graining of a wave field \cite{ASh2018},
which relaxes the resonance condition into $\mathbf{k}_0 +
\mathbf{k}_1 - \mathbf{k}_2 - \mathbf{k}_3 =
\Delta\mathbf{k}$.  In contrast to the standard condition
$\Delta\mathbf{k} = 0$,  the wavevector and frequency
mismatches satisfy $\Delta\omega/\omega_{min} <
\lambda_\omega$, $|\Delta \mathbf{k}| /k_{min} < \lambda_k
\bar \omega/\omega_{min}$, where $\omega_\mathrm{min}$ and
$k_\mathrm{min}$ are the minimum values of frequency and
wavenumber in the quartet, $\bar \omega$ is the mean
frequency, and $\lambda_\omega$ and $\lambda_k$ are the
detuning parameters. The crucial role is played by the
coarse-graining parameter $\lambda_k$. If $\lambda_k = 0$,
the wave field in the canonically transformed space, in
which both the Zakharov equation and the KE operate, is
free (gaussian) regardless of the amplitude, since a
logarithmically spaced grid with $\lambda_k = 0$ allows no
nontrivial wave interactions and, hence, no evolution. When
$\lambda_k$ is increased% %from zero
, the number of approximately resonant interactions (with
fixed $\lambda_\omega = 0.01$) grows approximately
quadratically with $\lambda_k$, while the rate of spectral
evolution, measured by the rate of change of various
integral parameters, quickly increases, until it reaches
saturation at a certain value $\lambda_k$, dependent on
grid resolution. Since in \cite{ASh2018} the value
$\lambda_k = 0.03$ was found to be optimal for the
$161\times 71$ grid, this value was used while computing
the DNS evolution shown in figure~1. For the purpose of
this study, it is convenient to use $\lambda_k$ as a way to
create wave fields with the same level of nonlinearity
$\varepsilon$, but different levels of non-gaussianity. In
order to avoid discreteness effects at low number of
interactions, we use the refined $321\times 71$ grid, along
with the $161\times 71$ one, and $0.003 \le \lambda_k \le
0.02$. Non-gaussianity can be measured as the ratio of the
nonlinear part of the ensemble averaged Hamiltonian $\bar
H$ \cite{AShWTBook}
\begin{equation}
\label{Hamb4avraged}
\bar H_{NL}  =   \displaystyle \frac 12\int T_{0123}
   \langle b_0^{*}b_1^{*}b_2b_3\rangle \delta _{0+1-2-3}\,{\rm d}{\mathbf k}_{0123},
\end{equation}
and its linear part $\bar H_{L} = \int \omega _0n_0\,{\rm
d}{\mathbf k}_0$.

\begin{figure} % figure 3
\includegraphics[width=8.6cm]{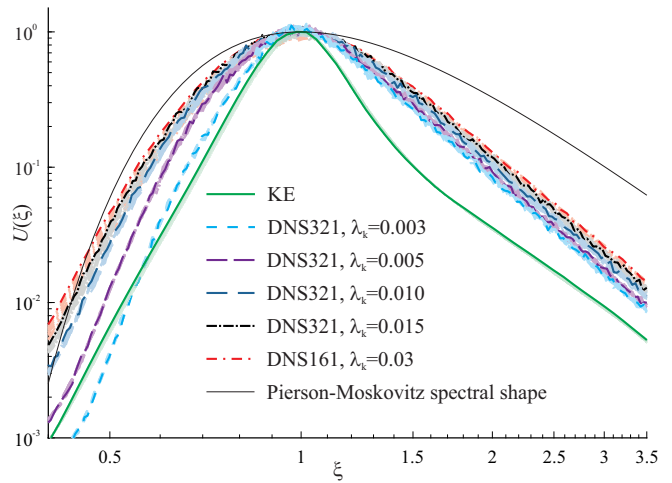}
 \caption{Self-similar shape function ${\cal U}(\xi)$, $\xi = k t^{6/11}$, extracted from the numerical solutions at the last 1000 wave periods of evolution. Shapes at every 100 periods are shown in light colours, the final curve is in darker colour of the same hue, normalized for
${\cal U}(1)=1$.
Numerical models used are the KE and the DNS on $161\times 71$ (DNS161) and $321\times 71$ (DNS321) grids for different values of $\lambda_k$
}
\end{figure}   % figure 3

Evolution of all wave fields is traced by the DNS with the
same wind forcing as above, until the peak of wave action
spectrum $n(k)$ reaches $k_p = 1$. Under the wind forcing,
linear and nonlinear parts of the Hamiltonian both grow
with time, but their ratio $\mu = \bar H_{NL}/\bar H_{L}$
is approximately constant at the self-similar stage of
evolution. The non-zero value of $\mu$, although very
small, is a prerequisite for spectral evolution. Figure 2
shows the value of $\mu$ averaged over the last 600
periods of evolution before reaching $k_p = 1$ for both
grids, as function of $\lambda_k$. Non-gaussianity quickly
grows with the increase of $\lambda_k$, approaching saturation
at $\lambda_k \ge 0.03$. Meanwhile,
the spectral evolution, as described by various integral
parameters, does converge for both grids, and is very close
between them. Formally, the value of $\lambda_k$ required
for the simulation can be made smaller by further
refinement of the grid, although in practice this is
limited by the available computational resources. In particular, 
on the refined $321\times 71$ grid the number of wave interactions
exceeds $10^9$ already for $\lambda_k = 0.01$.

We are mainly interested in the shape of the spectrum. At
the self-similar stage of evolution the spectral shape is
characterized by the self-similar function $\cal U$ for the
duration-limited evolution:
 $n(k) = a t^{23/11} {\cal U}(b k t^{6/11})$,
where $a$ and $b$ are constants \cite{Badulin2005}. The
simulated self-similar spectral shapes ${\cal U}(\xi)$ are
shown in figure 3. While the spectral slope 
depends on the parameterization of breaking more than on
$\lambda_k$, the shape of the spectral front
demonstrates a clear dependence on $\lambda_k$. For small
$\lambda_k$, the wave field has relatively few wave
interactions (about $10^8$ on the $321\times 71$ grid at
$\lambda_k = 0.003$), and the evolution is very slow,
although the self-similar state is eventually reached, with
the spectral front of the shape function ${\cal U}(\xi)$
close to the KE shape function. With the increase of
$\lambda_k$ from zero, the shape function quickly
approaches a different form, with a more rounded front and
wider peak, resembling the Pierson-Moskowitz spectral
shape.

\begin{figure} % figure 4
\includegraphics[width=8.6cm]{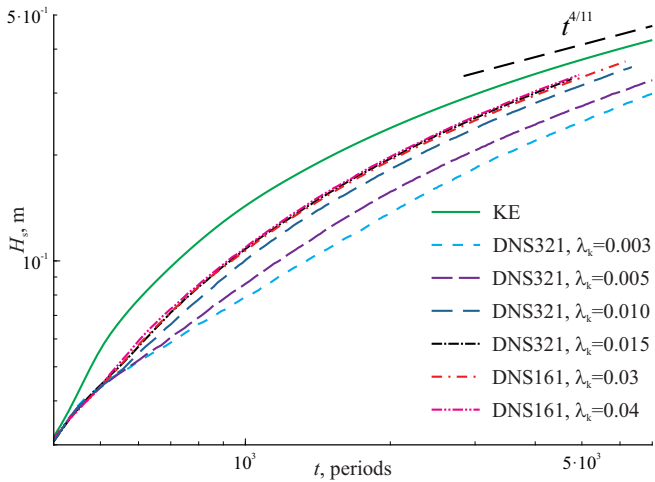}
\caption{Evolution of significant wave height $H_s$: DNS on $161\times 71$ and $321\times 71$ grids for different values of $\lambda_k$ vs the KE simulation}
\end{figure}   % figure 4

For many applications, the evolution and prediction of integral
charateristics of spectra, such as significant wave height,
is of particular importance \cite{Komen}.
%The theory based on the KE makes
%use of the homogeneity property of the right-hand side of
%the KE with respect to the spectrum, which leads to the
%universality of the wind wave growth rate, so that wind
%wave growth laws can be presented in a concise form that
%does not contain wind parameters \cite{Zakh2015}.
The important question is whether the evolution of integral
characteristics can be affected by the effects of finite
non-gaussianity for realistic wind speeds. To clarify this
point, we plot in figure 4 the evolution of significant
wave height $H_s$ obtained with the KE and the DNS for various
values of $\lambda_k$. Figure 4 demonstrates that with the
increase of $\lambda_k$ the DNS evolution of $H_s$
converges, remaining
slightly slower than that predicted by the KE, but
following the same asymptotic rate of increase known for the KE in the case
of constant action flux \cite{Badulin2005}. The difference
between the KE and the DNS is mostly manifested in the
spectral shape, while the difference in significant wave
height is relatively small and appears to be due to
breaking that effectively reduces the forcing. Simulation
of the KE with wind forcing reduced by 5\% makes the
difference insignificant.

\emph{Duscussion.}---Studies of wave kinetics based on the
KE rely on the homogeneity of the right-hand side of the
equation with respect to the spectrum. This property is an
artefact of the  neglect of non-gaussianity effects. In
this work, we show that although non-gaussianity is weak,
in the long term it leads to considerable distortion of the
spectral shape. At the same time, integral parameters of a
wave field appear to be much less affected, with the error
remaining within the uncertainty introduced by wave
breaking, which the DNS modelling has to take into account.
The spectral shape obtained by the DNS appears to be in
much closer agreement with observations of mature sea
states than the KE spectral shape.

The present study and its findings have numerous important
implications. First, they are of crucial importance for all
applications where the shape of the wave spectrum is
significant, rather than just its integrated description,
in particular for probability estimates of extreme wave
events, design or coastal hazard risk assessments, sediment
transport models, etc. Second, it is well known that the
wind wave models based on the KE are  optimized for certain
frequency and directional resolutions against the available
measurements. Knowledge of systematic errors in models can
drastically improve the quality of such optimizations, and
thus improve the quality of wind wave modelling. Third, the
findings of this study provide an insight on the role of
non-gaussianity in kinetic models, which is significant for
a wide context of wave turbulence in various branches of
physics.

\emph{Acknowledgments.}---The work was supported by UK NERC
grant NE/I01229X/1. Computations were performed on the CUDA
High Performance Computing cluster at Keele University and
on the ECMWF Supercomputing facility within the Special
Project SPGBSHRI.

\bibliography{Annenkov_Shrira_PRE}% Produces the bibliography via BibTeX.

\end{document}